\documentclass[conference]{IEEEtran}

\usepackage{cite}
\usepackage{amsmath,amssymb,amsfonts}
\usepackage{algorithmic}
\usepackage{graphicx}
\usepackage{textcomp}
\usepackage{xcolor}
\usepackage{comment}
\usepackage{listings}
\usepackage{url}

\def\BibTeX{{\rm B\kern-.05em{\sc i\kern-.025em b}\kern-.08em
    T\kern-.1667em\lower.7ex\hbox{E}\kern-.125emX}}

\newcommand{\fblas}{{\footnotesize F}BLAS\ }
\newcommand{\fblass}{{\footnotesize F}BLAS}

\newcommand{\etal}{\textit{et al}.}
\newcommand{\OUR}{our binary128 GEMM }
\newcommand{\Err}{$E_{\rm L1}$}
\newcommand{\MAD}{multiply-add\ }

\begin{document}
\title{Accelerating 128-bit Floating-Point Matrix Multiplication on FPGAs}

\author{
	\IEEEauthorblockN{Fumiya Kono\IEEEauthorrefmark{1}\IEEEauthorrefmark{4}, Naohito Nakasato\IEEEauthorrefmark{2}, Maho Nakata\IEEEauthorrefmark{3}}
 	\IEEEauthorblockA{\IEEEauthorrefmark{1} Shizuoka Institute of Science and Technology, Fukuroi, Shizuoka, JAPAN}
	\IEEEauthorblockA{\IEEEauthorrefmark{2} The University of Aizu, Aizuwakamatsu, Fukushima, JAPAN}
	\IEEEauthorblockA{\IEEEauthorrefmark{3} Cluster for Pioneering Research, RIKEN, Wako, Saitama, JAPAN}
	\IEEEauthorblockA{\IEEEauthorrefmark{4} kono.fumiya@sist.ac.jp}
}

\maketitle

\begin{abstract}
General Matrix Multiplication (GEMM) is a fundamental operation widely used in scientific computations. Its performance and accuracy significantly impact the performance and accuracy of applications that depend on it. One such application is semidefinite programming (SDP), and it often requires binary128 or higher precision arithmetic to solve problems involving SDP stably. However, only some processors support binary128 arithmetic, which makes SDP solvers generally slow. In this study, we focused on accelerating GEMM with binary128 arithmetic on field-programmable gate arrays (FPGAs) to enable the flexible design of accelerators for the desired computations. Our binary128 GEMM designs on a recent high-performance FPGA achieved approximately 90GFlops, 147x faster than the computation executed on a recent CPU with 20 threads for large matrices. Using our binary128 GEMM design on the FPGA, we successfully accelerated two numerical applications: LU decomposition and SDP problems, for the first time.
\end{abstract}


\IEEEpeerreviewmaketitle

\section{Introduction}
General Matrix Multiplication (GEMM) is a crucial computation in various scientific and engineering algorithms. Its precision plays a significant role in determining the accuracy of the target applications. Different applications have different precision requirements for the number of bits used to represent floating-point (FP) numbers. As defined by the IEEE 754 standard~\cite{ieee}, FP formats and arithmetic are available in various precisions, including binary16 (also known as half-precision), binary32 (single-precision), binary64 (double-precision), and binary128 (quadruple-precision). The suffix in each format indicates the number of FP bits supported by the respective format, with higher numbers indicating higher precision.

In machine learning (ML) using artificial neural networks, it has been shown that binary16 is sufficient for storing the weights of these networks. This has led to the development of hardware architectures that support highly parallel computation using binary16 arithmetic. One example is the TensorCore on recent NVIDIA graphics processing units (GPUs), designed for matrix multiplication with lower precision and has multiplication and accumulation performed in binary16 and binary32 arithmetics, respectively. Other ML accelerators, such as Google's TPUv3~\cite{9351692}, also support the bfloat16 format, an extended half-precision FP format.

On the other hand, operations with higher precision, such as binary128, are also required by specific applications. One example is Semidefinite Programming (SDP), a natural extension of linear programming that aims to minimize linear functions subject to certain constraints. In semidefinite programming, it is common to solve given problems using the Primal-Dual Interior-Point Methods (PDIPM)~\cite{doi:10.1137/1038003}. However, according to these methods, SDP is numerically unstable near the optimal solution because the variable matrices become singular~\cite{Alizadeh97,nakata}. Therefore, Nakata~\cite{nakata} proposed using higher precision numbers to solve optimization problems using SDP to maintain the desired numerical accuracy. 

However, since few processors represented by the IBM z13 processor~\cite{7563276} support binary128 as hardware, the performance of applications relying on binary128 arithmetic is typically 100 to 1000x slower than that only relying on binary64. Therefore, the acceleration of binary128 arithmetic is crucial for accelerating SDP. 

In this research, we implemented GEMM in binary128 arithmetic on Field Programmable Gate Arrays (FPGAs). The advantage of targeting FPGAs is their flexibility in optimizing accelerators for target computations. Additionally, while GPUs are designed with many parallel processors and fast memories, FPGAs are simply an array of logic gates that allow us to reconfigure designs and how they work during computation. This characteristic of FPGA enables us to create a suitable design for specific calculations while minimizing the use of hardware resources. As a result, energy consumption during computation on FPGAs is typically much lower than on GPUs.

Nagasu~\etal{}~\cite{nagasu} compared the energy consumption of FPGA and GPU computations for the same tsunami modeling application and demonstrated the effectiveness of FPGAs. They showed that their implementation on the Arria10 FPGA consumed approximately 5x less energy than the initial implementation on an AMD Radeon GPU.

Implementing logic designs on FPGAs is typically more challenging than parallel programming on GPUs because logic designs must be written in Hardware Description Language (HDL). To alleviate this difficulty, we adopt Intel's OpenCL-based high-level synthesis (HLS) techniques for \OUR designs in this research.

To design high-performance GEMM operations on FPGAs, it is essential to utilize pipeline parallelism and create a systolic array~\cite{KungSystolic}. Matteis~\etal{}~\cite{fblas} developed \fblass, a numerical library inspired by the open-source implementation of the Basic Linear Algebra Subroutines (BLAS) for Intel FPGAs. \fblas also provides a version of the systolic array design for its GEMM implementation.
In this research, we extended it to support various FP precisions.

The OpenCL standard supports neither arithmetic operations of binary128 nor that of higher precision than binary128. Furthermore, the OpenCL standard only supports arithmetic operations in binary32 and binary64~\cite{ieee}. While a recent version of the OpenCL SDK for Intel FPGAs supports specific FP precisions, its main target is binary16 and bfloat16 for machine learning.

In this research, we adopted customized FP units developed by Nakasato~\etal{}~\cite{nakasato} that support various FP formats, including the binary128 format. Nevertheless, this paper focused on developing and evaluating binary128 format FP addition, multiplication units, and acceleration of binary128 GEMM operations.

The main contributions of this research are as follows:

\begin{itemize}
\item We implemented fast GEMM designs in the binary128 format on FPGAs
\item We developed an application interface compatible with the standard BLAS library.
\item We evaluated the performance of \OUR designs with practical applications.
\end{itemize}

While this research builds upon the preceding work, we successfully integrated \OUR designs into MPLAPACK~\cite{mplapack}, an extension of all BLAS and LAPACK (Linear Algebra PACKage) routines to support multi-precision FP operations, including binary128. Therefore, the designs can also be immediately used in numerical applications that utilize MPLAPACK as a backend.

Our binary128 GEMM design implemented on Terasic DE10a-Net Agilex FPGA achieved  90.9GFlops by utilizing maximum hardware resources. Furthermore, its integration to practical applications of blocked LU decomposition and SDP contributed to at most 5.3x and 2x speed-up compared with the computation on a recent Intel i9-10900 CPU with 20 threads parallelization by OpenMP, respectively.

This paper first presents a brief specification of \OUR designs. Then, to inspect the fundamental characteristics of the designs, we first evaluate their performance on Terasic DE5a-Net Arria10 FPGA. Based on the analysis obtained by this evaluation, we focus on more practical benchmarking by using Nallatech (BittWare) 520N Stratix10 FPGA, which is installed on a supercomputer system in operation, and Agilex FPGA, the latest and high-end Intel FPGA. Finally, we discuss the applications of \OUR design by integrating it to blocked LU decomposition and SDP problems.

\section{Related Works}
The study of GEMM in high-precision arithmetic is a popular topic in multiple-precision research, but previous studies have mainly focused on CPU or GPU implementations.

Nakasato~\cite{nakasato2} accelerated the GEMM routines for binary32, binary64, and 128-bit double-double (DD)~\cite{Dekker_1971,Kunuth_1998} precision on the AMD Cypress GPU. 
Also, Nakata~\etal{}~\cite{nakata2} presented a fast GEMM implementation in DD precision on NVIDIA GPUs. In the paper, they have applied their GEMM implementation in DD precision to the algorithm in SDP. 
Kouya~\cite{kouya} implemented LU decomposition supporting multi-precision floating-point numbers such as DD, triple-double (TD), and quad-double (QD). With AVX vectorization, the implementation successfully accelerated the LU decomposition for Intel and AMD CPUs.

Joldes~\etal{}~\cite{campary} developed CAMPARY, a multi-precision arithmetic library for NVIDIA GPUs based on the CUDA programming model, which supports DD, TD, and QD precision. 
Isupov and Knyazkov have been working on MPRES-BLAS for NVIDIA GPUs~\cite{10.1007/978-3-030-64616-5_4}, which is an interval evaluation for the fractional representation of numbers in the Residue Number System (RNS)~\cite{doi:10.1137/1011027} to represent arbitrary precision numbers. MPRES-BLAS was the fastest among CAMPARY and CUMP~\cite{CUMP} GEMM implementation for 424-bit precision. 

Mukunoki~\etal{}~\cite{mukunoki} also had proposed a fast GEMM implementation in binary128 or less precision based on the Ozaki scheme~\cite{ozaki}, an accurate GEMM algorithm by representing FP numbers as non-overlapping sums of FP numbers.
They showed the performance evaluation of their method on CPUs and prospects of extension for GPUs.

However, research for GEMM in high-precision arithmetic on FPGA has yet to be seen. Licht~\etal{}~\cite{Licht} targeted Xilinx FPGAs to implement GEMM by using systolic array designs. Afterward, they experimented with their GEMM to support various FP precision up to 1024-bits~\cite{Licht2} extending the implementation of the Multiple Precision Floating-Point Reliable (MPFR)~\cite{mpfr} library.
Although their motivation lies in the acceleration of an SDP solver,
the practical evaluation of their designs still needs to be done.

\section{Matrix Multiplication for FPGA}
\subsection{Implementation}
The GEMM routine in BLAS performs matrix multiplication for matrices $A$ and $B$ as follows:

\begin{equation}
C = \alpha A B + \beta C,
\label{GEMM_DEF}
\end{equation}
where $\alpha$ and $\beta$ are scalar parameters.
Listing~\ref{fort_dgemm} presents the API in C language to the GEMM routine for multi-precision FP numbers called {\it Rgemm} provided by MPLAPACK \cite{mplapack}. Note that \_Float128 is the standard data type in C language for binary128, as defined in ISO/IEC TS 18661-3:2015 \cite{ISOIECTS1866132015}. MPLAPACK utilizes \_Float128 through the GNU C++ compiler via GNU extensions.
The first two arguments specify the transpose operation of matrices $A$ and $B$. The three arguments {\it lda}, {\it ldb}, and {\it ldc} represent the leading dimensions of matrices $A$, $B$, and $C$, respectively.

\begin{figure*}
\begin{lstlisting}[caption = RGEMM interface, label = fort_dgemm]
void Rgemm(const char *transa, const char *transb, 
            int const m, int const n, int const k, _Float128 const alpha, 
            _Float128 *a, int const lda, _Float128 *b, int const ldb, 
            _Float128 const beta, _Float128 *c, int const ldc);
\end{lstlisting}
\end{figure*}

In the practical implementation of the GEMM routine, calculating the matrix multiplication $AB$ is a critical part of its computation. Assume that we have two matrices $A$ and $B$ with sizes $m \times k$ and $k \times n$, respectively. Then, an element of the resulting matrix $C' = AB$ is computed by the summation as follows:
\begin{equation}
	C'_{ij} = \sum_{p=0}^{k-1} A_{ip}\times B_{pj},
	\label{mm}
\end{equation}
where $i,j$, and $p$ are indices ranging $0 \leq i < m$, $0 \leq j < n$,  and  $0 \leq p < k$, respectively. The calculation of the whole matrix $C'$ involves a 3-level nested loop.

\begin{figure}[h]
\centerline{\includegraphics[width=90mm]{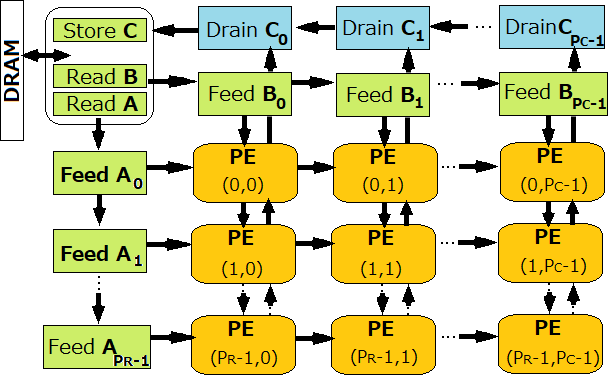}}
\caption{Systolic Array Design for the GEMM operation}
\label{systolic}
\end{figure}

Fig.~\ref{systolic} illustrates the design of a systolic array for \OUR design derived from \fblas \cite{fblas}. This design is characterized by a 2-D array of processing elements (PE) aligned $P_C \times P_R$.
Each PE calculates Eq.~(\ref{mm}) for assigned sub-matrices of $A$ and $B$.
The size of the sub-matrices $A$ and $B$ and the value of $P_C \times P_R$ determine how the input matrices are partitioned.

In the computation flow, the input matrices $A$ and $B$ are read from main memory via the {\tt Read} module and sent to the PEs through the {\tt Feed} module. $A$ is sent by column, and $B$ is sent by row, assuming that both matrices are not transposed. They are first received by PEs with IDs $(P_R-1, 0)$ or $(0, P_C-1)$ and forwarded to the adjacent PEs in the systolic array on each clock cycle. Each PE accumulates the result of a \MAD operation for the same element in $C'$ and sends it to the {\tt Drain} module, which is eventually collected by the {\tt Store} module to be written back to the main memory.

More specifically, \fblas is a generator of OpenCL kernels for the systolic array. The generated systolic array consists of four OpenCL kernels: two kernels that combine the {\tt Read} and {\tt Feed} modules for $A$ and $B$, one {\tt Store} kernel for $C$, and a main kernel for the array of PEs and {\tt Drain} module.
The main kernel explicitly calls a function for one PE in a loop. By fully unrolling the loop, the main kernel defines the systolic array. Because the computation task of a PE is just a \MAD operation, we can replace the \MAD operation in the original design with any \MAD unit for a desired FP format. This enables us to create a systolic array design corresponding to the designated precision.

In addition, to replace the \MAD operation, we modify and extend the other three kernels for the {\tt Read}, {\tt Feed}, and {\tt Store} modules to support a wider memory bus for binary128 arithmetic. We also extend the original kernels to optimize load and store operations from DRAM. The {\tt Read} and {\tt Feed} kernels are equipped with a memory buffer in front of the {\tt Feed} module. In the original design, the memory buffer is called a {\tt memory tile} and explicitly instantiated as a 1-D array. The {\tt memory tile} acts as a cache memory to store a sub-matrix of $A$ and reuse the sub-matrix many times. 
The exploitation of the {\tt memory tile} reduces the pressure on the memory bandwidth of DRAM and improves the performance of \OUR designs, as shown in the later section.


The number of PEs in the present systolic array is $P_R \times P_C$. We instantiate $P_R \times P_C$ binary128 \MAD units. The additional computations in the definition of the GEMM, as shown in Eq.~(\ref{GEMM_DEF}),
require the computation of two scaler-matrix multiplications and one matrix addition, which are very costly in a GEMM design on an FPGA. In the present systolic array, we need additional $P_C$ multiply units for $\alpha A$, a load unit for $C$, $P_C$ multiply units for $\beta C$, and $P_C$ add units for the summation of $\alpha A$ and $\beta C$. Except for the multiply units for $\alpha A$, which can be merged with the {\tt Feed} module, the other units are only activated in the final stage of the GEMM operation at the {\tt Store} module. Therefore, in this research, we only calculate Eq.~(\ref{mm}) on an FPGA, while the host CPU handles the transpose operations and other additional operations involving $\alpha$ and $\beta$. Supporting those additional operations, we develop an API that is compatible with the standard {\it Rgemm} provided by MPLAPACK.
It enables us to use \OUR designs immediately in numerical applications with minimal changes. 

\subsection{Performance Models}
Here, we summarize the performance models for \OUR design. In this section, $f$ represents the clock frequency of the logic circuit design in MHz.

\subsubsection{Performance of GEMM}
The peak performance of the designs depends on the layout of systolic arrays, as shown in Fig.~\ref{systolic}. When we use $P_R \times P_C$ PEs, the peak performance $F_{\rm peak}$ (GFlops) is given by Eq.(\ref{peak}).

\begin{equation}
	F_{\rm peak} = \dfrac{2\times P_R \times P_C \times f \times 10^6}{10^9}
	\label{peak}
\end{equation}

The measured performance $F_{\rm perf}$ of the designs in GFlops is calculated by Eq.~(\ref{flops}), where $T_{\rm exec}$ is the execution time in seconds.

\begin{equation}
	F_{\rm perf} = \dfrac{2mnk}{T_{\rm exec} \times 10^9}
	\label{flops}
\end{equation}

In Eq.~(\ref{mm}), $m, n$ and $k$ denote the matrix size parameters. For the multiplication of $n \times n$ square matrices, the number of FP operations is $2n^3$.

\subsubsection{Memory Bandwidth Requirement}

The performance of the designs is also affected by memory bandwidth of an FPGA board. $P_R \times P_C$ systolic array takes $P_R + P_C$ inputs conveyed by two vertical and horizontal {\tt Feed} pipelines at every cycle. Thus, the required memory bandwidth $B_{\rm req}$ (GB/s) is given by Eq.~(\ref{MBreq}).

\begin{equation}
	B_{\rm req} = \dfrac{(P_R + P_C) \times f \times 10^6 \times N_{\rm Byte}}{10^9}
	\label{MBreq}
\end{equation}

$N_{\rm Byte}$ represents the word size established as 16 bytes in the present work. If the systolic array consists of $8 \times 8$ PEs, $B_{\rm req}$
equals $256f \times 10^{-3}$ GB/s. For example, the requirement $B_{\rm req}$ becomes 51.2GB/s for the design where the clock frequency $f$ is 200MHz. To fully utilize all PEs in the designs, $B_{\rm req}$ must be smaller than the memory bandwidth of a target FPGA board.

\section{Performance Evaluation}
This section presents the performance evaluation of \OUR designs on three FPGA systems.

\subsection{Benchmarking Conditions}

\subsubsection{Target FPGA Systems}

Table~\ref{spec} shows the specification of FPGAs used in this benchmarking: Terasic DE5a-Net Arria10, Nallatech (BittWare) 520N Stratix10, and Terasic DE10a-Net Agilex. The Stratix10 FPGA is a computation node of Cygnus, a supercomputer system operating at the University of Tsukuba in Japan since 2019. We use Intel FPGA SDK for OpenCL to design and implement \OUR designs. A different host system hosts each FPGA as specified in the bottom rows of Table~\ref{spec}.

\begin{table*}[h]
\begin{center}
	\caption{Specification of FPGA systems in our performance evaluation}
		\vspace{-2mm}
		\small
		\begin{tabular}{c||c|c|c}
			FPGA & Arria10 & Stratix10 & Agilex \\ 
			\hline \hline
			Logic cells & 427,200  &  933,120 &  487,200  \\ 
			\hline
			DSP blocks & 1,518 & 5,760 & 4,510 \\ 
			\hline
			M20K RAM blocks & 2,713 & 11,721& 7,110 \\ 
			\hline
			Memory bits (total) & 55,562,240 & 240,046,080 & 145,612,800\\ 
			\hline
			Board Memory & $2\times$ DDR3-1066 8GB  & $4\times$ DDR4-2400 8GB & $4\times$ DDR4-2666 8GB  \\ 
            \hline
			Board Memory Bandwidth & 34.2GB/s  & 76.8GB/s & 85.2GB/s \\ 
            \hline
			PCIe & Gen2 x8  & Gen3 x8 & Gen3 x16 \\ 
			\hline
			Quartus ver. & 19.1 & 20.4 & 21.1\\
   
			\hline \hline
                        CPU & i7-2600K & Xeon Gold 6226 & i9-10900\\
                        \hline
                        $N_{\rm core}$/$N_{\rm thread}$ & 4/8 & 16/32 & 10/20\\
                        \hline
                            Host memory size & 16GB &  192 GB & 64 GB\\
			\hline
			Host OS & Ubuntu 18.04.6LTS & CentOS 7.9.2009  &  Ubuntu 20.04.5LTS  \\ 
			\hline
			{\it gcc} ver. & 8.4.0 & 7.4.0 & 9.4.0 
                            
                        \label{spec}
		\end{tabular}
	\end{center}
\end{table*}

\subsubsection{Evaluation Method}
We first evaluate \OUR designs for square matrices by scaling $n$.
Also, we evaluate the performance of multiplying non-square matrices with sizes $m \times k$ and $k \times n$ as more realistic and practical evaluations. 
To calculate the performance in GFlops, Eqs.~(\ref{peak}) and (\ref{flops}) are used. The computation time $T_{\rm exec}$ in Eq.~(\ref{flops}) is the average of three trials in each benchmarking. As a target of comparison, we use a baseline of the {\it Rgemm} executed on the host system of Agilex (i9-10900 CPU) with 20 threads by OpenMP parallelization.

Besides, we compare numerical accuracy with the {\it Rgemm} routine provided by MPLAPACK on a CPU. As shown in Eq.~(\ref{errs}), we calculate the average L1 norm of the difference between two $n \times n$ matrices as \Err\  throughout the evaluation.

\begin{equation}
	E_{\rm L1} = \dfrac{\displaystyle \sum_{i=0}^{n-1}\sum_{j=0}^{n-1} \left| C^F_{ij} - C^R_{ij} \right| }{n^2},
	\label{errs}
\end{equation}
In Eq.~(\ref{errs}), $C^F$ and $C^R$ denote the result matrices by our implementation for FPGAs and {\it Rgemm}, respectively.
\Err\ allows us to determine how accurately \OUR designs match the results of the reference implementation.

To highlight the main characteristics of computational performance, we begin by evaluating the designs on the Arria10 FPGA in this section. The following section covers the performance evaluation of the designs on newer FPGAs, including Stratix10 and Agilex.

\subsection{Benchmarking Results on Arria10}
\subsubsection{Evaluation for Square Matrices}

\label{gemm_perf}
We present benchmarking results for \OUR designs. The systolic array consists of PEs arranged in a square with $P_R=P_C=2,4$, and 8. Table~\ref{a10synth} shows the logic synthesis results on the Arria10 FPGA system.
\begin{table*}[h]
	\begin{center}
		\caption{Synthesis results of each $P_R \times P_C$ designs on Arria10 FPGA}
		\small
		\begin{tabular}{c||c|c|c}
			$P_R \times P_C$  & $2\times 2$ & $4\times 4$ & $8\times 8$ \\ 
    \hline
        $M_{{\tt Tile}}$ & 32 & 32 & 32 \\
			\hline\hline
			Logic cells&  49,523 (12\%) & 78,624 (18\%) & 201,033 (47\%)\\
			\hline
			DSP blocks &  78 (5\%) & 270  (18\%)  & 1,037 (68\%)  \\
			\hline
			Memory bits& 3,390,704 (6\%) & 3,841,264 (7\%)  & 5,341,616 (10\%) \\
			\hline
			RAM blocks & 397  (15\%) & 514 (19\%) & 551 (20\%)  \\	
			\hline
			Fmax (MHz) &  236.29 & 228.15 &  201.28 \\ 
			\hline
			$F_{\rm peak}$ (GFlops) &  1.89  & 7.30 &  25.76 \\ 
		\end{tabular}
		\label{a10synth}
	\end{center}
\end{table*}

\begin{table*}[h]
\begin{center}
	\caption{Synthesis results of each $P_R \times P_C$ designs on Stratix10 and Agilex FPGAs}
	\small
	\begin{tabular}{c||c|c||c|c}
            FPGA & \multicolumn{2}{|c||}{Stratix10} & \multicolumn{2}{c}{Agilex}\\
            \hline\hline
		$P_R \times P_C$  & $8\times 8$ & $8\times 16$  & $8\times 8$  & $8\times 16$ \\ \hline
        $M_{{\tt Tile}}$ & 128 & 256 & 128 & 512\\
		\hline\hline
		Logic cells& 354,019 (38\%)  & 524,014 (56\%)  & 288,693 (59\%) & 414,416 (85\%)\\
		\hline
		DSP blocks & 1,038 (18\%)        & 2,062  (36\%)      & 1,293 (29\%)  &  2,061 (46\%)\\
		\hline
		Memory bits& 18,294,816 (8\%) & 32,191,072 (13\%) & 18,969,348 (13\%) & 79,467,844 (55\%)\\
		\hline
		RAM blocks & 1,292  (11\%)      & 1,792 (15\%)   & 1,510 (21\%)  & 4,289 (60\%)\\	
		\hline
		Fmax (MHz) &  259.06         &  177.14         &  411.52  & 388.95\\ 
		\hline
		$F_{\rm peak}$ (GFlops) &  33.16  & 45.35 &  52.67 & 99.57\\ 
	\end{tabular}
	\label{s10agi_synth}
\end{center}
\end{table*}

Our binary128 GEMM design requires more DSP blocks for larger PE arrays.
Therefore, the number of available DSP blocks is the primary constraint for the design. The row labeled {\rm Fmax} shows the clock frequency of each design. Therefore, their peak performance $F_{\rm peak}$ is shown in the last row based on Eq.~(\ref{peak}).

\begin{figure}[h]
	\centerline{\includegraphics[width=90mm]{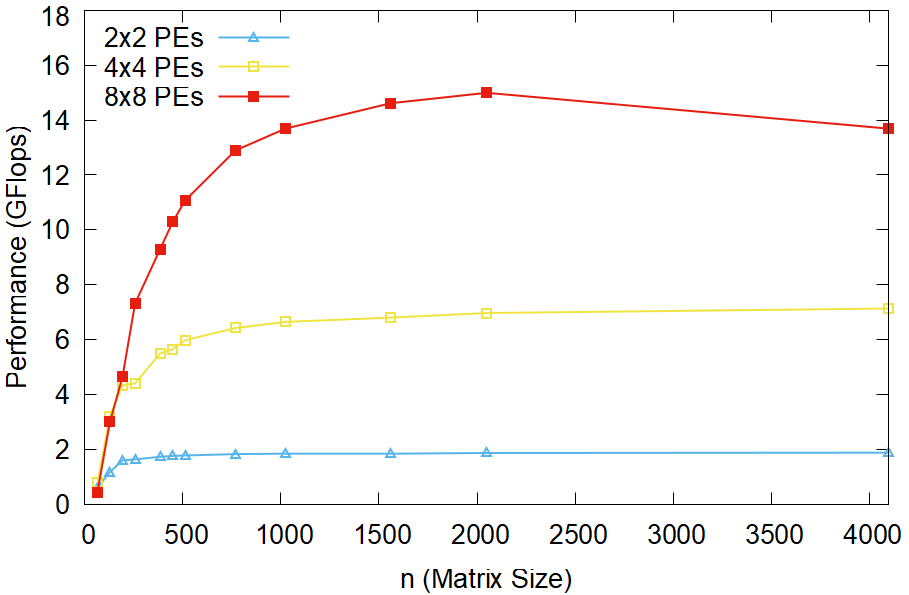}}        
        \vspace{-3mm}
	\caption{Performance of \OUR designs for square matrices on Arria10 FPGA}
	\label{gemm_a10}
\end{figure}

Fig.~\ref{gemm_a10} shows the performance of each design on Arria10. The matrix size $n$ ranges from 64 to 4096. The performance of designs $F_{\rm perf}$ with $2\times 2$, $4\times 4$, and $8\times 8$ PEs is at a maximum of 1.88, 7.1, and 15.0GFlops, respectively. Since each PE can work independently for data streaming and operations on the systolic array, the performance is proportional to the number of PEs in the design.

However, with a small $n$, the computation load for each PE is not sufficiently high to reach the maximum performance of the designs. It reaches the peak at a specific $n$,  such as $n=2048$ for $8\times 8$ PEs, and the performance scaling becomes flat at larger $n$.

We then evaluate the numerical error \Err\ of computation results between \OUR designs and the {\it Rgemm} routine based on Eq.~(\ref{errs}). \Err\ for $n<512$ is distributed between $10^{-31}$ and $10^{-30}$. As we set $n$ to 4096, \Err\ increases to $2.0\times 10^{-28}$. The layout of PEs does not make a significant difference in \Err.

Regarding the comparison between $F_{\rm perf}$ and $F_{\rm peak}$,
a ratio to designs of $2\times 2$, $4\times 4$, and $8\times 8$ PEs is 99.5\%, 97.3\%, and 58.2\%, respectively. 
Recall that the memory bandwidth requirement $B_{\rm req}$ is given by Eq.~(\ref{MBreq}). As we substitute {\rm Fmax} of each design in Fig.~\ref{gemm_a10} with $f$ in Eq.~(\ref{MBreq}), we find $B_{\rm req}$ 15.1GB/s, 29.2GB/s and 51.5GB/s for $2\times 2$, $4\times 4$ and $8\times 8$, respectively. 

Our Arria10 system has two DDR3 memories that provide 34.2GB/s of the total bandwidth.
It is sufficient for the designs of $2\times 2$ and $4\times 4$ PEs. As a result, their $F_{\rm perf}$ is close to the peak. However, the design of $8\times 8$ PEs requires 51.5GB/s, which is 1.5x more significant than the available bandwidth.
Therefore, the design of $8\times 8$ PEs is limited by memory transfer from DRAM. 
As a result, we see that the ratio between $F_{\rm perf}$ and $F_{\rm peak}$ is much lower than that of other designs of fewer PEs.

\subsubsection{Effects of Memory Buffer for The Systolic Array}
To enhance performance, we instantiate more PEs in \OUR design. However, the memory bandwidth of the FPGA board poses a limitation. Therefore, the systolic array generated by \fblas has a module called {\tt memory tile} in front of the {\tt Feed} module. It is a local memory buffer working as a cache memory for each PE to mitigate the memory bandwidth requirements provided in Eq.~(\ref{MBreq}). As the systolic array incorporates a more significant number of PEs, increasing the size of $M_{{\tt Tile}}$ is necessary to provide the larger buffer in \OUR designs.

\begin{figure}[h]
	\centerline{\includegraphics[width=90mm]{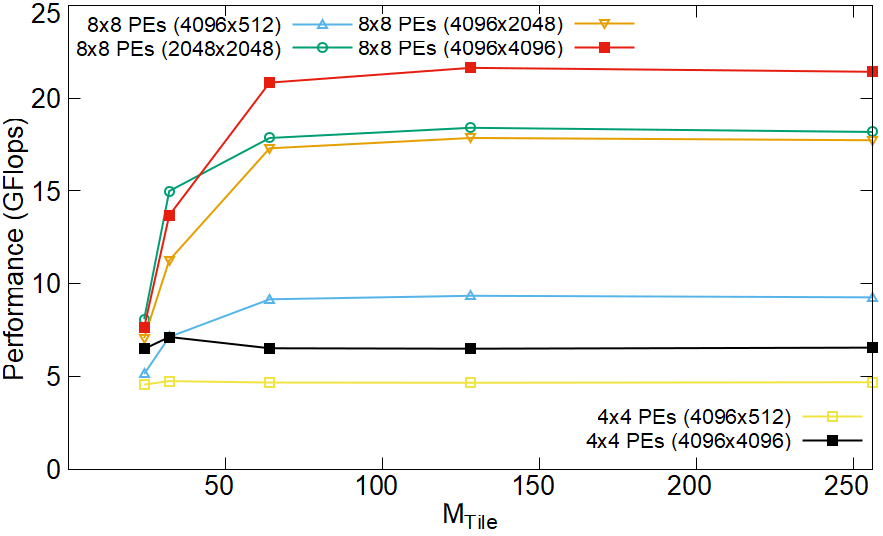}}
	\caption{Performance of \OUR designs on Arria10 FPGA with $M_{{\tt Tile}}=$ 24 to 256}
	\label{gemm_a10tile}
\end{figure}

The results presented in Sec.~\ref{gemm_perf} were all obtained by the designs with $M_{{\tt Tile}}=32$. We then conduct additional benchmarking to further investigate the potential performance improvement by adopting a larger value of $M_{{\tt Tile}}$.
Fig.~\ref{gemm_a10tile} illustrates the performance of the GEMM by using the designs of $4\times 4$ and $8\times 8$ PEs where $M_{{\tt Tile}}$ ranges from 24 to 256.

The figure shows the performance of each design for four matrices where $(k,n)=$ $(4096,512), (4096,2048)$, $(2048,2048)$, $(4096,4096)$ assuming $m=k$. Computations using the design of $4\times 4$ PEs are not affected by the change of $M_{{\tt Tile}}$ since their $B_{\tt req}$ (30.25GB/s) is within the board memory bandwidth (34.2GB/s).

On the other hand, we see that using a larger $M_{{\tt Tile}} \geq 64$ improves the performance of the $8\times 8$ PEs. In those cases, the performance increases by 1.5 to 2x compared to the design with $M_{{\tt Tile}}=32$ and reaches its peak at $M_{{\tt Tile}}=128$. In contrast, the smaller $M_{{\tt Tile}}\leq 24$ causes even lower performance. For the square matrix with $n=4096$, we achieved 21.6GFlops at $M_{{\tt Tile}}=128$, 84\% of $F_{\rm peak}$ in Table~\ref{a10synth}. We also see that this $M_{{\tt Tile}}$ scaling is effective in multiplying tall-skinny matrices where $n$ is relatively much smaller than $k$. The larger $M_{{\tt Tile}}$ reduces a bottleneck of the current implementation to some extent.

\subsubsection{Evaluation for Non-square matrix}
\begin{figure}[h]
	\centerline{\includegraphics[width=90mm]{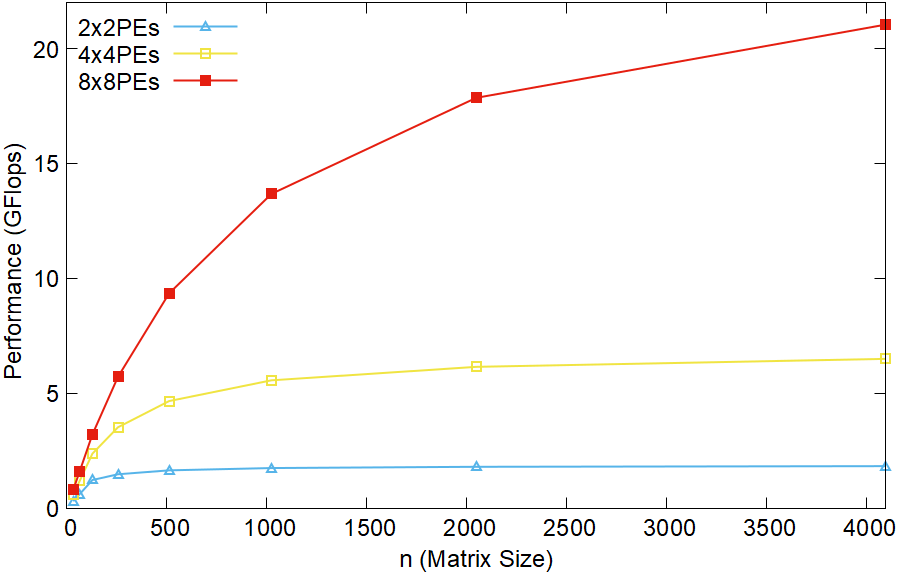}}
         \vspace{-3mm}
	\caption{Performance of \OUR designs on Arria10 FPGA for non-square matrices where $n$ ranges from 32 to 4096}
	\label{gemm_a10r}
\end{figure}

In computation of square matrices, we found that the performance of \OUR designs was ideal, except for the memory bandwidth constraint caused by the large PE layout. We then evaluate the performance for non-square matrices. Fig.~\ref{gemm_a10r} shows the result gained by multiplications of $m \times k$ and $k \times n$ matrices where $m$ and $k$ are fixed at $m=k=4096$ and only $n$ is varied between 32 and 4096. In this evaluation, we set $M_{{\tt Tile}}=128$ in all designs. 

In the case of multiplication with rectangular matrices, the current systolic array design is ineffective due to load imbalance among PEs.
However, when the layout of PEs is small, such as $2\times 2$ PEs, the performance does not drop even for multiplication with $4096 \times 128$ compared to $4096 \times 4096$. 

However, the multiplication on the design of $8\times 8$ PEs clearly shows a performance degradation for any $n$. In particular, for the computations of tall-skinny matrices where $n$ is much smaller than $k$, the design of $8\times 8$ PEs performs far from its maximum capacity. The performance is as low as that of $2\times 2$ PEs. When we similarly fix $m$ and $n$ to $m=n=4096$ and scale $k$ between 32 to 4096, the computation of each design shows the same result as in Fig.~\ref{gemm_a10r}.

\subsection{Benchmarking Results on Stratix10 and Agilex}

We then evaluate \OUR designs on Stratix10 and Agilex FPGAs under the same benchmarking conditions.
Based on the previous evaluation of Arria10, the designs targeted in this section are $8\times 8$ PEs with $M_{{\tt Tile}}=128$. Additionally, we implemented a design of $8\times 16$ PEs with $M_{{\tt Tile}}=256$ and 512 to utilize the abundant hardware resources on Stratix10 and Agilex.
However, their resources are still insufficient to implement $16\times 16$ PEs due to the limited number of available logic cells.

Table~\ref{s10agi_synth} summarizes the logic synthesis results of our designs implemented on each FPGA.
As we increase the size of the memory buffer on each PE by scaling $M_{{\tt Tile}}$,
the utilization of memory bits and RAM blocks on the FPGAs accordingly increases.
However, this does not cause problems on the Stratix10 and Agilex FPGA systems when we set $M_{{\tt Tile}}=512$ for $8\times 16$ PEs. As a result, {\rm Fmax} and $F_{\rm peak}$ for \OUR designs on Stratix10 and Agilex are much higher than those on Arria10.

\begin{figure}[h]
	\centerline{\includegraphics[width=90mm]{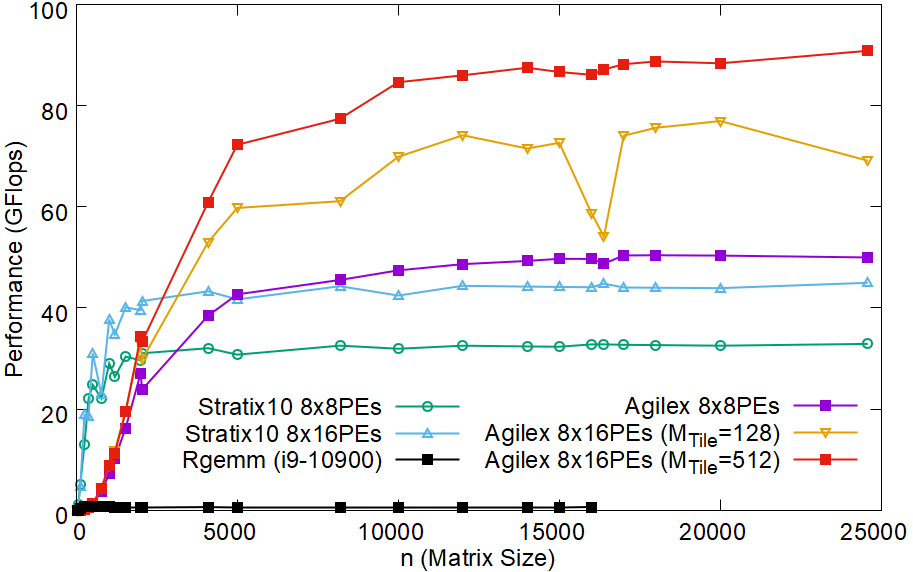}}
         \vspace{-3mm}
	\caption{Performance of \OUR designs for square matrices on Stratix10 and Agilex FPGAs}
	\label{gemm_s10agilex}
\end{figure}

Fig.~\ref{gemm_s10agilex} shows the performance of \OUR designs on the two FPGAs. On FPGA systems of Stratix10 and Agilex,
we could execute GEMM with the size of a maximum $n=24576$ thanks to their large board memory. For comparison, we plot the performance on a host CPU (i9-10900) in the Agilex FPGA system.

We first focus on results for Stratix10. The design of $8\times 8$ PEs with $M_{{\tt Tile}}=128$ almost reached its peak performance at $n=4096$. The performance scaling for larger $n$ is at 32.8GFlops, 99\% of the peak. $8\times 16$ PEs with $M_{{\tt Tile}}=256$ similarly reached a peak of 45.0GFlops at around $n=12000$. However, compared to the design of $8\times 8$ PEs, its performance improvement is sluggish because the {\rm Fmax} of the $8\times 16$ PEs significantly dropped and led to a low $F_{\rm peak}$ of the design.

As we examine the performance of the designs on Agilex, the optimization of PE layout and $M_{{\tt Tile}}$
successfully contributed to performance improvement. 
While the design of $8\times 8$ PEs with $M_{{\tt Tile}}=128$ certainly performs effectively, that of $8\times 16$ PEs with $M_{{\tt Tile}}=512$ is much better. The computation by the $8\times 16$ PEs achieved 90.9GFlops, 91\% of the peak, for the largest matrix size of $n=24576$
in contrast to one by the $8\times 8$ PEs yielding 50.4GFlops at $n=18000$, about 96\% of its peak.

The importance of the size of $M_{{\tt Tile}}$ can be easily understood by comparing it
with a reference plot for the design of $8\times 16$ PEs with $M_{{\tt Tile}}=128$ on Agilex. 
If we set $M_{{\tt Tile}}=128$, the performance of the design is at most 77GFlops, which is only 77\% of the peak. In particular, a trench in the plot at $n=16384$ results in a significant performance drop to 54.1GFlops around that point.
One reason may be that those specific large matrices accidentally cause accesses that stride over different memory banks on four independent DIMMs on the Agilex FPGA board. However, the memory buffer exploited by the larger $M_{{\tt Tile}}$ (e.g. 512) helps to alleviate problems related to unexpected memory access patterns and facilitates steady performance improvement.

Finally, \OUR design is very high performance compared to the {\it Rgemm} routine executed on the CPU with 20 threads. Its performance settles at 650MFlops for $n>1024$. Therefore, we have a significant advantage in processing large matrices. The design of $8\times 16$ PEs with $M_{{\tt Tile}}=512$ on Agilex is 145x faster than the computation on a recent CPU with the maximum number of threads.

\begin{figure}[h]
	\centerline{\includegraphics[width=90mm]{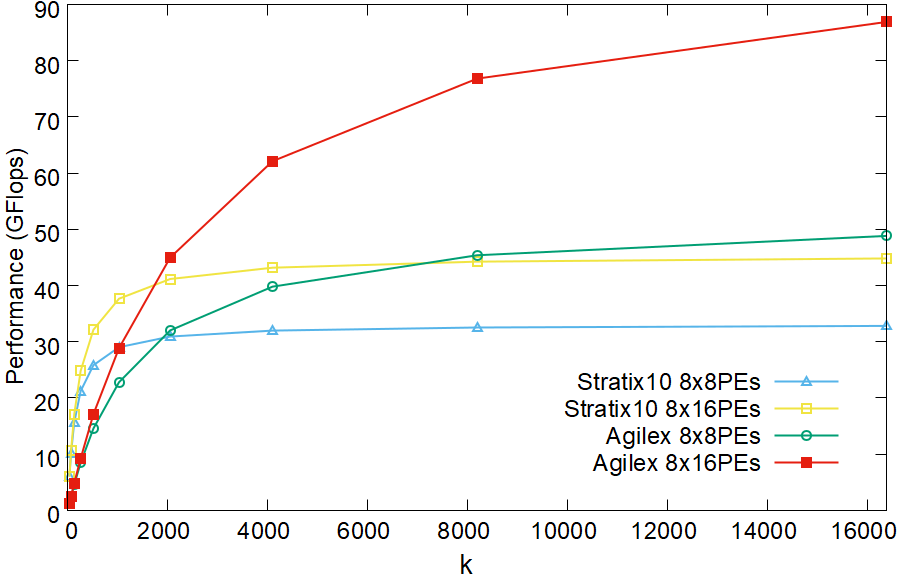}}
         \vspace{-3mm}
	\caption{Performance of \OUR designs on Stratix10 and Agilex FPGAs for non-square matrices where $k$ ranges from 32 to 16384}
	\label{gemm_s10agilex_rect}
\end{figure}

In addition, we show the performance of \OUR designs for non-square matrices on Stratix and Agilex FPGAs. Fig.~\ref{gemm_s10agilex_rect} shows the benchmarking result when $m$ and $n$ are fixed to $m=n=16384$, and $k$ is scaled between 32 and 16384. As presented in the benchmarking on Arria10, the performance drop for ratios of $n:k<2:1$ is not significant. However, for tall-skinny matrices where $k$ is particularly small, like $k\leq 128$, even the performance on Agilex is just a few GFlops. As a result, the advantage of \OUR designs compared to computation on CPUs is lost.

\section{Application of binary128 Matrix Multiplication}
Once we have \OUR designs by the systolic array architecture, we can accelerate practical applications which require binary128 GEMM operations. We here describe two applications of our implementation with performance evaluation. In this section, $\mathbb{R}^{n\times n}$ denotes $n\times n$ real matrices.

\subsection{Blocked LU Decomposition}
\label{BLU}
\subsubsection{Problem Specification of LU Decomposition}
The LU Decomposition is a fundamental operation in numerical analysis that factorizes the given square matrix $A$ as a multiplication of lower and upper triangular matrices like $A = LU$ where $L$ and $U$ are lower and upper triangular matrices, respectively.
Based on BLAS routines, the LU decomposition in binary64 precision is implemented as a routine called {\it dgetrf} in LAPACK. 
The {\it degetf} routine adopts a blocked LU decomposition algorithm thoroughly investigated and implemented for every supercomputer in the last four decades. Its variation is the most famous parallel benchmarking program called LINPACK.
The blocked LU decomposition algorithm effectively solves dense linear equations on accelerator architectures like GPU since its computation is mainly processed as GEMM operations.

\begin{figure}[h]
\vspace{-3mm}
        \centerline{\includegraphics[width=50mm]{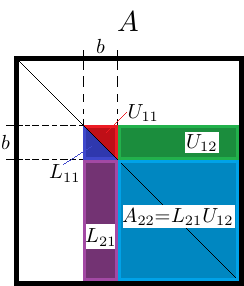}}
	\caption{Blocked LU Decomposition of a matrix $A$ where the block size is $b$}
	\label{lu}
\end{figure}

Let us consider the LU decomposition for a matrix $A \in \mathbb{R}^{n\times n}$ with the block size $b$,
as shown in Fig.~\ref{lu}. Then, we obtain $L$ and $U$ on $A$ by repeating the following procedure recursively.

\begin{enumerate}
	\item Divide $A$ into 4 sub-matrices: $A_{11} \in \mathbb{R}^{b\times b}$, $A_{12} \in \mathbb{R}^{b\times (n-b)}$, $A_{21} \in \mathbb{R}^{(n-b)\times n}$, and $A_{22} \in \mathbb{R}^{(n-b)\times (n-b)}$.
	\item Perform decomposition $A_{11} = L_{11}U_{11}$.
	\item Solve $U_{12}$ that satisfies $L_{11}U_{12} = A_{12}$.
	\item Solve $L_{21}$ that satisfies $L_{21}U_{11} = A_{21}$.
	\item Update $A_{22}$ by $A_{22} = A_{22} - L_{21}U_{12}$.
	\item If $n-b>0$ still holds, go back to step 1 after substituting $A$ with $A_{22}$.
\end{enumerate}


In step 5, we have matrix multiplication $L_{21}U_{12}$. When $b = 1$, the blocked LU decomposition is reduced to a non-blocked routine called {\it dgetrf2} in LAPACK. When $b$ is large enough, the computation of {\it dgetrf} is dominated by GEMM operations in step 5. Accordingly, it can be accelerated by GEMM routines on GPUs or FPGAs. 

In MPLAPACK \cite{mplapack}, all BLAS and LAPACK routines are extended to support multi-precision FP operations, including binary128. We modify an extended version of {\it dgetrf} in MPLAPACK called {\it Rgetrf}, which calls the {\it Rgemm} routine. In this paper, we replace calls to {\it Rgemm}
with \OUR operations executed on FPGAs.

The number of FP operations in the LU decomposition algorithm is $\dfrac{2n^3}{3} - \dfrac{n^2}{2} + \dfrac{5n}{6}$ \cite{lapack_note}. Here, we regard it as $\dfrac{2n^3}{3}$. Therefore, $F'_{\rm perf}$ as shown in Eq.~(\ref{flops_lu}) gives the computation performance for the following evaluation.

\begin{equation}
	F'_{\rm perf} = \dfrac{2n^3}{3\times T_{\rm exec} \times 10^9}
	\label{flops_lu}
\end{equation}

\subsubsection{Evaluation of GEMM for LU Decomposition}

We assume that an input $n \times n$ matrices whose elements are given by random numbers in a range of $[0.0, 1.0)$. Then, the input matrices can be factorized by the LU decomposition. We decompose the square matrices by applying \OUR designs in the algorithm. 

Based on the evaluation in the previous section, we measure the performance of blocked LU decomposition with the design of $8 \times 16$ PEs on Agilex FPGA. We scale the size of matrices $n$ and apply different block sizes $b$ to find the optimal size of $b$. As a comparison, we present a result on the design of $8 \times 16$ PEs on Stratix10 where $b = 128$. We also give an another comparison with a result obtained through computation using only the host CPU (Intel Core i9-10900). In that computation, the {\it Rgetrf} routine in MPLAPACK takes charge of the LU decomposition with 20 threads by OpenMP parallelization.

\begin{figure}[h]
  	\centerline{\includegraphics[width=90mm]{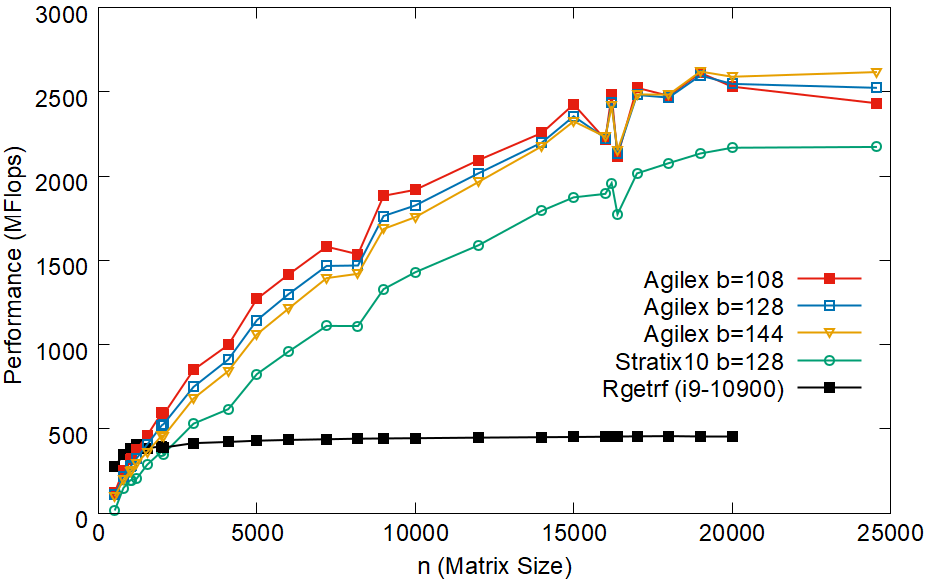}}
        \vspace{-3mm}
	\caption{Performance of LU decomposition on Stratix10 and Agilex FPGAs}
	\label{lu_s10agi}
\end{figure}

Fig.~\ref{lu_s10agi} summarizes our results of the LU decomposition. For Agilex FPGA,
we present the performance in each case of $b=108, 128, 144$.
The black line shows the performance scaling obtained by the computation on the CPU.

We observe that $b = 108$ yields the best performance on the Agilex FPGA as represented by 2.5GFlops at $n=20000$.
However, with a large matrix of $n = 24576$, a higher $b$ yields the peak. We can see in the figure that the highest performance is 2.6GFlops obtained with $b=144$ for the matrix of $n=24576$. On the other hand, the performance deteriorates when we apply even larger values of $b$ such as $b=192$ and $256$, yielding 2.3GFlops and 2.1GFlops, respectively. Similarly, the design on the Stratix10 FPGA is superior to the CPU computation for $n>3000$. Although it is slower than the computation on the Agilex FPGA, it finally reaches 2.2GFlops at $n=20000$, which is 4.7x faster than that of the CPU.

Since the performance on FPGAs improves slowly by scaling $n$ until computation data saturate every PE, the performance on the CPU for small $n$ is superior to that of FPGAs. When the matrix size $n = 512$, the smallest size in this evaluation, the performance on the CPU is 278MFlops which is 2 to 3x faster than that of FPGAs.
We see that the intersection of the performance scaling between the CPU and FPGAs is around $n = 1536$. The performance of the CPU execution does not improve for $n > 2000$, which is 458MFlops at $n=24576$. In contrast, the performance of the LU decomposition by using \OUR designs on Agilex FPGA is at a maximum of 5.3x faster than that of the CPU.

We compare the decomposed matrices $L$ and $U$ calculated by the designs on FPGAs with the reference result calculated by the CPU by using Eq.~(\ref{errs}). In the case of $n\leq 1536$, where the CPU computation is still faster than FPGAs, we find \Err\ $\sim 10^{-31}$.
On the other hand, as we test for the matrix of $n=24576$,
we find \Err\ $\sim 10^{-28}$. This consequence is the same as we expected, considering the previous evaluation of \OUR design.

Finally, we compare our results with those of previous work by Kouya \cite{kouya},
who presented optimizations of LU decomposition using DD arithmetic. 
Specifically, they have applied memory blocking and vectorization using AVX2 instructions
and evaluated the performance on an Intel Core i9-10900X CPU. According to their benchmarking for $n = 1024$, the performance of a conventional blocked LU decomposition code with $b=64$ was 132MFlops. Similarly, the performance of a vectorized LU decomposition code with $b=32$ was 363MFlops. In contrast, our result with the design of $8\times 16$ PEs achieved 324.5MFlops for $n=1024$ and $b=108$ on an Agilex FPGA. Even the fastest design on the high-end FPGA is not significantly beneficial for small matrices.
As a result, from performance perspective for small matrices,
\OUR designs are inferior to the vectorized LU decomposition code on a CPU.

However, we emphasize that our designs on recent FPGAs are much more effective for large $n$.
With the current best performance of our LU decomposition being 2.5GFlops, our FPGA designs are superior for large matrices. It is also worth noting that our work and the work by Kouya \cite{kouya} use different FP formats.
DD arithmetic is well suited for recent high-end CPUs equipped with vector arithmetic units such as AVX2 and AVX512 instructions on the x86-64 ISA, Neon, and SVE instructions on the ARM ISA.

\subsection{Semidefinite Programming (SDP)}

SDP is an optimization problem to minimize or maximize a given linear function under
the constraint of symmetric semidefinite matrices. It has vast possible applications in engineering~\cite{Ohsaki}, finance~\cite{Adrian}, quantum chemistry~\cite{Fukuda}, and physics~\cite{RevModPhys.91.015002}, which have been investigated for a long time.

SDPA\cite{Yamashita} is a numerical implementation and software package for SDP written in C++ \cite{sdpa}.
The algorithm used in the SDPA is called the PDIPM, one of the iteration methods for SDP.
Previous research \cite{nakata} has extended the SDPA to support various precision
FP operations such as SDPA-GMP, -DD, and QD~\cite{nakata}. The GMP version uses arbitrary precision arithmetic.
Thus, a user must specify the precision beforehand. These extended versions of the SDPA use a part of MPLAPACK\cite{mplapack} as a back-end, mainly through calling the {\it Rgemm} routine. 

To determine which parameters are utilized in GEMM routines called from the SDPA, we conduct 92 problems provided by SDPLIB\cite{sdplib} using SDPA-binary128 with MPLAPACK. As we are currently focusing on accelerating GEMM routines in our work, we have modified the code to record the 13 arguments specified in Listing \ref{fort_dgemm} for the {\it Rgemm} routine during the execution of all problems. 

Analysis of the collected data reveals that the SDPA frequently calls the {\it Rgemm} routine with non-square matrices,
and none of the leading dimensions of the matrices in the {\it Rgemm} routine equal $m$, $n$, or $k$. Of the over 800 combinations of arguments recorded in the collected data, we find only 50 combinations where the condition $n = m = k = lda = ldb = ldc$ holds. As shown in Sec.~\ref{gemm_perf}, the performance of \OUR designs on FPGAs for non-square matrices is inferior to that for square matrices.

Based on our analysis, we evaluate the performance of the SDPA calling
{\it Rgemm} operation accelerated by an FPGA
only when either two conditions are satisfied; (1) $m$ equals $n$ or
(2) $m \times n \times k $ is larger than a predefined parameter $N_{\rm min} = 10^6$.
We test different $N_{\rm min}$ and find that $N_{\rm min} = 10^6 $ to $ 10^7$ is optimal for the SDPA.
We only present the performance benchmarking of the SDPA on Agilex FPGA 
for selected problems from SDPLIB shown in Table~\ref{sdpa_table}.
We present the elapsed time per iteration of the SDPA-binary128 
on the three systems: CPU-A (Intel Xeon Gold 5122 4 cores @ 3.60GHz), 
CPU-B (Intel i9-10900 CPU 10 cores @ 2.80GHz),  and CPU-B using \OUR design of $8 \times 16$ PEs on Agilex.
The performance with the FPGA is 2 to 4x and roughly 1.5x faster than that of CPU-A and CPU-B, respectively.
Note that the performance of SDPA-binary128 on CPUs is proportional to the number of cores on a given CPU.

We verify that each solution computed by \OUR design improves upon the solution obtained via double-precision calculations.
As illustrated in Table~\ref{sdpa_table2}, we present the relative gaps, primal/dual feasible errors, and the numbers of iterations for problems theta2, theta3, theta4, theta6, and controll11 from SDPLIB,
as computed on CPU-B using binary128, FPGA (Agilex) using our design,
the DD precision version~\cite{nakata}, and the double precision version~\cite{Yamashita}.
As smaller errors indicate better results, the solutions obtained via \OUR design exhibit
an improvement over those obtained via double precision calculations and are of comparable or slightly superior quality to those obtained via DD arithmetic.
Our binary128 {\it Rgemm} accelerated by FPGAs effectively accelerates the PDIPM for SDP problems.

\begin{table}[h]	
\begin{center}
\caption{Elapsed time per iteration in sec. of the SDPA on CPUs and Agilex FPGA}
\vspace{-2mm}
\small
\begin{tabular}{c||c|c|c}
Problem  & CPU-A & CPU-B & FPGA(Agilex) \\
\hline
theta2    & 0.8     & 0.45    & 0.42    \\
theta3    & 4.99    & 2.68    & 2.11    \\
theta4    & 21.17   & 10.24   & 7.28    \\
theta5    & 69.35   & 30.82   & 20.17   \\
theta6    & 191.4   & 79.54   & 48.3    \\
control11 & 66.92   & 38.09   & 28.51   \\
equalG51  & 141.04  & 66.87   & 33.32   \\
gpp500-1  & 18.45   & 8.53    & 4.47    \\
gpp500-2  & 18.58   & 8.53    & 4.56    \\
maxG11    & 53.35   & 25.66   & 16.39   \\
maxG32    & 803.42  & 380.92  & 232.81  \\
maxG51    & 108.69  & 53.34   & 34.19   \\
mpc500-1  & 11.49   & 5.39    & 3.36    \\
mpc500-4  & 14.37   & 7.31    & 4.9     \\
qpG11     & 264.96  & 111.95  & 64.48   \\
qpG51     & 480.79  & 207.78  & 120.98  \\
thetaG11  & 82.11   & 41.7    & 28.55   \\
thetaG51  & 853.03  & 387.67  & 248.86  
\label{sdpa_table}
\end{tabular}
\end{center}
\end{table}

\begin{table}[h]	
\begin{center}
\caption{The relative gaps, primal/dual feasible errors, and the number of iterations for certain problems from SDPLIB were calculated on CPU-B using binary128, FPGA (Agilex, binary128), the double-double precision version (DD), and the double precision version.}
\small
\begin{tabular}{l||r|r|r|r}
Problem  & CPU-B & FPGA  & DD  & double \\
 \hline
 theta2 & & & & \\
 relative gap      &  1.05e-24     & 1.16e-24    &  2.68e-25     &  1.45e-08     \\
 p.feas.error      &  3.70e-32     & 7.70e-33    &  4.93e-31     &  3.55e-15     \\
 d.feas.error      &  2.14e-25     & 2.89e-25    &  4.51e-27     &  5.77e-15     \\
\# of iterations   & 58  & 62& 51  & 17  \\ \hline
 theta3 & & & & \\
 relative gap      &  5.71e-25     & 5.35e-24    &  1.86e-23     &  1.57e-08     \\
 p.feas.error      &  1.08e-32     & 1.23e-32    &  9.86e-31     &  8.88e-15     \\
 d.feas.error      &  2.43e-26     & 6.40e-25    &  1.42e-24     &  4.00e-15     \\
\# of iterations   & 55  & 50& 61  & 17  \\ \hline
 theta4 & & & & \\
 relative gap      &  5.15e-25     & 5.89e-25    &  6.18e-27     &  2.25e-08     \\
 p.feas.error      &  2.62e-32     & 1.85e-32    &  7.89e-31     &  7.11e-15     \\
 d.feas.error      &  3.51e-26     & 9.35e-26    &  5.06e-28     &  1.47e-14     \\
\# of iterations   & 71  & 94& 52  & 18  \\ \hline
 theta6 & & & & \\
 relative gap      &  6.90e-31     & 1.23e-30    &  6.28e-25     &  2.45e-08     \\
 p.feas.error      &  1.39e-32     & 1.85e-32    &  8.87e-31     &  1.42e-14     \\
 d.feas.error      &  5.74e-32     & 5.80e-32    &  5.19e-26     &  5.04e-14     \\
\# of iterations   & 45  & 46& 54  & 18  \\ \hline
 control11 & & & & \\
 relative gap      &  9.01e-25     & 1.87e-23    &  2.24e-22     &  8.26e-06     \\
 p.feas.error      &  1.62e-27     & 2.02e-27    &  2.41e-25     &  1.86e-09     \\
 d.feas.error      &  1.60e-24     & 4.51e-24    &  1.50e-22     &  2.03e-07     \\
\# of iterations   & 64  & 62& 60  & 47  \\ \hline
\end{tabular}
\label{sdpa_table2}
\end{center}
\end{table}

\subsection{Discussions on Application Performance}
The blocked LU decomposition algorithm {\it Rgetrf} outlined in Sec.~\ref{BLU}
employs the {\it Rgemm} operation to compute $A_{22} = L_{21}U_{12}$,
where both matrices are non-square and skinny.
$L_{21}$ and $U_{12}$ are matrices of dimensions $b \times k$ and $k \times b$, respectively.
During the loop from step 2 to step 6, $k$ is reduced as $k = n - pb$,
where $p$ represents the iteration number starting from $p = 1$.
At an initial phase of the algorithm, $k$ is large enough such that \OUR designs on the Agilex FPGA
effectively accelerate the performance of {\it Rgetrf}.
However, as $k$ becomes much smaller than $n$ at a later phase of the algorithm,
the acceleration by the Agilex FPGA becomes ineffective.
The blocking size $b$ also impacts the performance of the GEMM on FPGAs.
For instance, if $b$ is too small, the performance of {\it Rgemm} on FPGAs
is significantly reduced, as depicted in Figs.~\ref{gemm_a10} and \ref{gemm_s10agilex}.

On the other hand, the PDIPM frequently calls the {\it Rgemm} operation
for small non-square matrices with a wide range of combinations of matrix sizes $n$, $k$, and $m$.
The largest matrix size in all problems presented in Table \ref{sdpa_table} is only $n = k = m = 2000$.
With a matrix size of $n = k = m = 2000$, the performance of {\it Rgemm} on FPGAs is half the peak performance.
In most cases, the algorithm calls the {\it Rgemm} operation for much smaller matrices
when it is not executed on the FPGA.
In a previous evaluation of a fast GEMM in DD arithmetic on GPUs by Nakata~\etal{} ~\cite{nakata2},
it was shown that the performance of the PDIPM in DD arithmetic accelerated by a GPU
is more than 10x faster than that on a CPU with four cores.
According to their results, the size of matrices does not significantly affect the performance
of {\it Rgemm} on GPU. Therefore, they have always utilized the GPU, except for very small matrices.

Despite the superior performance of our accelerated {\it Rgemm} implementation on the Agilex FPGA,
which is more than 100x faster than the reference {\it Rgemm} on a 10-core CPU,
the two applications evaluated in this section are not substantially accelerated by the FPGA.
Therefore, to make \OUR designs on FPGAs more practical for real-world applications,
we will need to extensively modify the systolic array design generated by \fblas
to address the performance degradation for small matrices and non-square matrices.
A potential solution is to develop an extended version of {\it Rgemm} that incorporates another level of blocking in the host code. Specifically, we could develop a new {\it Rgemm} API based on a batched GEMM algorithm \cite{10.1145/2716282.2716288}.  
It would allow us to instantiate multiple systolic arrays on an FPGA to handle the batched GEMM algorithm.
One of a hardware implementation of a batched GEMM algorithm focusing on 64 and smaller bits of FP numbers was reported by Ledoux~\etal{} \cite{batchedGEMM}. Their systolic array design leverages a stalling-free output scheme for the output matrix $C$ to maximize the overlap of host data transfers with GEMM computations.

\section{Conclusion}

In this paper, we presented \OUR implementation and its evaluation of different Intel FPGAs, and
its integration into numerical applications such as blocked LU decomposition and SDP.
Our GEMM designs on FPGAs are based on the 2-D systolic array generated by the \fblas library.
Furthermore, by optimizing memory buffer size, which stores reused data in fast on-chip memory,
we successfully implemented $8\times 16$ PEs to accelerate the GEMM in
binary128 arithmetic on FPGAs.

The benchmarking in this paper showed that our implementation is particularly advantageous when computing large matrices of size $n>10^4$.
For example, in our evaluation of \OUR implementation on the Agilex FPGA, the performance was 90.9GFlops, 91\% of the estimated peak performance of the design. This resulted in a 147x speed-up compared to the {\it Rgemm} routine provided by MPLAPACK on an i9-10900 CPU with 20 threads.

Further benchmarking of various matrix multiplications showed that our designs are pretty effective to accelerate
GEMM operations for square and almost-square matrices. In other words, LU decomposition can be solved faster using our implementation than with existing CPU routines. However, our design was not effective at handling tall-skinny matrices, commonly found to solve semidefinite programming.

Our current systolic array designs for GEMM operations are based on the OpenCL kernels
generated by the latest version of \fblas \cite{fblas_github}.
The \fblas is designed to be flexible and accommodate various kernel configurations
for different BLAS routines, such as General Matrix-Vector Multiplication (GEMV) and
Triangular Solve with Multiple Right-Hand Sides (TRSM).
However, in this study, we extracted only the systolic array kernels of GEMM for our work.
Extending our work to other BLAS routines would be an interesting area for future research.

There is still room for optimization to improve the performance of our GEMM design
when we use it to calculate tall-skinny matrix multiplications.
Further optimizations are necessary to achieve the desired performance, especially for SDP problems.
In future work, we will compare such optimized GEMM designs with other high-precision GEMM
implementations on accelerators. Another area of future work will be to
explore other FP formats in our GEMM designs by replacing
the current binary128 \MAD units with \MAD units in different arithmetic.

\section*{Acknowledgment}
A part of this paper is based on results obtained from a project, JPNP16007, commissioned by the New Energy and Industrial Technology Development Organization (NEDO). This work was partly supported by MEXT as ”Feasibility studies for the next-generation computing infrastructure” and KAKENHI Grant Number JP23K11133. 

This research in part used computational resources of Cygnus provided by Multidisciplinary Cooperative Research Program in Center for Computational Sciences, University of Tsukuba.

We thank Prof. Ishikawa, High Energy Accelerator Research Organization, and Prof. Daisaka, Hitotsubashi University, Japan, for their help evaluating our designs on Stratix10.

\bibliographystyle{IEEEtran}
\bibliography{gemm}

\end{document}